\documentclass[letterpaper,twoside]{article}

\newcommand{\ttbar}     {$t\overline{t}$\ }
\newcommand{\MET}       {\mbox{$\not\!\!E_T$}\ }
\newcommand{\pt}        {$p_T$\ }
\usepackage{amssymb,amsmath}
\usepackage{hyperref}

\begin{document}

\title{Searches for New Physics in Top Decays at D0}



\author{Marc-Andr\'e Pleier\\Brookhaven National Laboratory, Physics Department,\\ Omega Group, Building 510A, Upton, NY  11973\\~\\
On behalf of the D0 Collaboration}
\date{}

\maketitle

\begin{abstract}
  The Tevatron proton-antiproton collider at Fermilab with its centre
  of mass energy of 1.96~TeV allows for pair production of top quarks
  and the study of top quark decay properties. This report reflects
  the current status of measurements of the W boson helicity in top quark 
  decays and the ratio of top quark branching fractions as well
  as searches for neutral current top quark decays and pair production
  of fourth generation t' quarks, performed by the D0 Collaboration
  utilising datasets of up to 5.4 fb$^{-1}$.
\end{abstract}


\section{Introduction}
With a mass of $m_t =$ 173.2 $\pm$ 0.9~GeV~\cite{topmass}, the top
quark is the most massive fundamental particle known to date.  Top
quarks are mainly produced in pairs via the strong interaction, with a
cross section of
$\sigma_{t\bar{t}}=7.3^{+0.5}_{-0.6}$~pb~\cite{topxsec} in $p\bar{p}$
collisions at $\sqrt{s}=1.96$~TeV. In the framework of the Standard
Model (SM), top quarks decay almost exclusively into $W$ bosons and
$b$ quarks.  Consequently, top quark pair events contain a $b$ and a
$\bar{b}$ quark from the \ttbar decay, and the remaining final state
to be observed depends on the decay mode of the two $W$ bosons.

The analyses presented here focus on two signatures: (i) events where
both $W$ bosons decay leptonically, resulting in a final state
containing two isolated high-\pt leptons, missing transverse energy
\MET corresponding to the two neutrinos and two ($b$-) jets ({\em dilepton}
events) and (ii) events where one $W$ boson decays leptonically, the
other one hadronically, resulting in one isolated high-\pt lepton,
\MET and four jets ({\em lepton + jets} events). As leptons only electrons 
and muons are considered in these analysis channels, including those
from leptonic $\tau$ decays.

\section{Searches for New Physics in Top Quark Pair Decays}
\subsection{Measurement of the $\mathbf{W}$ Boson Helicity}
Top quark decays in the SM proceed via the left-handed charged current
weak interaction, which results in a suppression of right-handed $W$
bosons. The expected fractions of the three possible helicity states
of the $W$ boson are: longitudinal fraction $f_0 \approx 69\%$,
left-handed fraction $f_- \approx 31\%$ and right-handed fraction $f_+
\approx 10^{-3}$~\cite{WHeltheo}. The $W$ boson helicity is reflected in the
angular distribution $\cos\theta^{\ast}$ of its decay products, with
$\theta^{\ast}$ being the angle of the down-type decay products of the
$W$ boson (charged lepton, $d$- or $s$-quark) in the $W$ boson rest
frame relative to the top quark direction. Any observation of $f_+$ at
the percent level would signal new physics beyond the SM.

D0 measures the $W$ boson helicity using a dataset corresponding to an
integrated luminosity of 5.4 fb$^{-1}$, using both lepton + jets and
dilepton final states of \ttbar decays~\cite{WHel}. Since the
measurement is still expected to be limited by sample statistics, the
event selection in each channel is optimised for \ttbar acceptance,
using multivariate likelihood discriminants based on
event topology and object kinematics. In the lepton + jets channel,
the event kinematics is obtained from the best kinematic fit to the 
\ttbar event hypothesis, which enables the calculation of 
$\cos\theta^{\ast}$ for the leptonic $W$ boson decay. The hadronic $W$
boson decay is utilised as well to evaluate $|\cos\theta^{\ast}|$,
which does not permit discrimination between right- and left-handed
$W$ bosons, but still adds information to further constrain $f_0$.
In the dilepton events, $m_t = 172.5$~GeV is assumed and $\cos\theta^{\ast}$
is obtained for each top quark as the average over all possible solutions
of the event kinematics, taking lepton and jet energy resolutions
into account by varying the energies within their uncertainties.

By comparing the observed angular distributions in data with templates
of background and \ttbar signal with varying $W$ boson helicity
admixtures, $f_0$ and $f_+$ can be extracted in a simultaneous fit
($f_- = 1 - f_+ - f_0$). This model-independent fit yields:
$$
f_{0} = 0.669 \pm 0.078\:{\rm(stat.)} \pm 0.065\:{\rm(syst.)},\:
f_{+} = 0.023 \pm 0.041\:{\rm(stat.)} \pm 0.034\:{\rm(syst.)},
$$
in excellent agreement with the SM expectation.

\subsection{Search for Neutral Current Top Decays}
Transitions between quarks of different flavour but identical charge
(flavour changing neutral currents, FCNC) are forbidden at lowest
order and heavily suppressed at higher orders in the SM. The decay
$t\to Zq$ (with $q = u,c$) thus is expected to occur with a branching
fraction of ${\mathcal O}(10^{-14})$~\cite{FCNCtheo}, well out of reach of
experimental sensitivity. Any observation of FCNC decays would
indicate physics beyond the SM.

D0 performs a search for \ttbar$\to WbZq+X$ decays based on a dataset
corresponding to an integrated luminosity of 4.1~fb$^{-1}$, using the
trilepton final state $\ell'\nu\ell\bar{\ell}$ + jets for the first
time in this kind of search~\cite{FCNC}. This signature benefits from
low background contributions, but also suffers from limited statistics.
After a basic preselection reflecting the decay signature, signal and
background (mainly diboson and $V$ + jets production) are separated
using distributions of jet multiplicity, scalar sum of the transverse
momenta of leptons, jets and \MET ($H_T$), and the reconstructed top
quark mass from the $t\to Zq$ decay.

All observed distributions are consistent with the SM background
contributions, and consequently 95\% C.L. limits on FCNC decays
of \ttbar are derived:
$$ {\mathcal B}(t\to Zq) < 3.2 \%, $$
with an expected limit of 3.8 \%. This is currently the world's best
limit on $ {\mathcal B}(t\to Zq)$.

\subsection{Search for Fourth Generation $\mathbf{t'}$ Quark Pair Production}
While a fourth generation of fermions with a light neutrino ($m_\nu <
m_Z / 2$) has been ruled out by electroweak precision data, the
constraints on a fourth generation are much relaxed if the neutrino is
sufficiently heavy. For example, a fourth generation $t'$ quark mass
of 400 GeV can be compatible with electroweak precision
measurements, and the mass splitting relative to the corresponding $b'$
quark is constrained to be small, leading to a preferred decay $t' \to
Wq$ ($q=d,s,b$)~\cite{tprimetheo}.

Based on a dataset with an integrated luminosity of 5.3 fb$^{-1}$, D0
searches for pair production of fourth generation up-type quarks
($t'\bar{t}'$), assuming the intrinsic $t'$ width is smaller than the
detector resolution and the decay always proceeds via
$Wq$~\cite{tprime}. Consequently, the $t'\bar{t}'$ decay chain is
identical to that of the top quark pair production, and the lepton +
jets final state is used for the search with no $b$-jet identification
applied. To separate the signal from SM background processes, the
observed events are compared with signal and background contributions
in two-dimensional histograms of the scalar sum of lepton-$p_T$,
jet-$p_T$ and \MET ($H_T$) versus $t'$ mass $m_{fit}$ from a kinematic
fit to the event hypothesis. Based on the observed agreement between
data and the SM contributions, 95\% C.L. limits are derived on the
$t'\bar{t}'$ production cross section as a function of $m_{t'}$.

While in the $e$ + jets channel reasonable agreement between data and
simulation is observed, in the $\mu$ + jets channel an excess of data
over SM background is visible around $m_{fit} = 325$~GeV. However, the
data are best described with a $t'\bar{t}'$ production cross section
$3.2 \pm 1.1$ times the predicted cross section for $m_{t'} =
325$~GeV. The probability of observing an excess of at least this size
from the SM background processes alone corresponds to 2.5 standard
deviations. Combining the observations from both channels, $t'\bar{t}'$
production can be excluded at 95\% C.L. for $$m_{t'} < 285~{\rm
GeV},$$ with an expected exclusion below 320~GeV.

\subsection{Measurement of $\mathbf{{\mathcal B}(t \rightarrow Wb) / {\mathcal B}(t \rightarrow Wq)}$}
The ratio of top quark branching fractions $R$ can be expressed via
CKM matrix elements as $R = $ \mbox{${\mathcal{B}}(t \rightarrow Wb)/$}
\mbox{$\Sigma_{q=d,s,b} {\mathcal{B}}(t \rightarrow Wq)$}$ = |V_{tb}|^{2}/$ 
\mbox{$\Sigma_{q=d,s,b} |V_{tq}|^{2}$} and provides therefore a measure
of the relative size of $|V_{tb}|$ compared to $|V_{td}|$ and
$|V_{ts}|$.  Within the framework of the SM, assuming a unitary
3$\times$3 CKM matrix and insignificance of non-$W$ boson decay modes
of the top quark, the above expression simplifies to $R = |V_{tb}|^{2}$
and $R$ is strongly constrained from global CKM fits:
$R=0.99830^{+00006}_{-00009}$~\cite{PDG2010}.  Deviations of $R$ from
unity could for example be a sign for the presence of a fourth heavy
quark generation or non-SM top quark decay modes.

D0 measures $R$ in a dataset corresponding to an integrated luminosity
of 5.4~fb$^{-1}$, using both the lepton + jets and dilepton \ttbar
decay channels~\cite{R}. In the lepton + jets channels, the number of
identified $b$-jets (0, 1, $\geq$2) is used together with a
multivariate kinematic discriminant to separate the \ttbar signal from
background in subsamples dominated by background in order to extract
$R$ simultaneously with the \ttbar production cross section
$\sigma_{t\bar{t}}$, yielding $R =0.95\pm 0.07{\rm~(stat.+syst.)}$ and
$\sigma_{t\bar{t}}= 7.90^{+0.79}_{-0.67}~{\rm (stat.+syst.)~pb}$.  In
the dilepton channels, the lowest value of a neural network $b$-jet
identification algorithm applied to the leading jets in each event is
used as a discriminant, yielding $R =0.86\pm 0.05{\rm~(stat.+syst.)}$
and $\sigma_{t\bar{t}}= 8.19^{+1.06}_{-0.92}~{\rm (stat.+syst.)~pb}$.
These measurements allow the extraction of $\sigma_{t\bar{t}}$ without
assuming ${\mathcal B}(t\to Wb)= 1$ as usually done in \ttbar
cross section measurements. Combining both measurements yields: $$R
=0.90\pm 0.04{\rm~(stat.+syst.)~and~} \sigma_{t\bar{t}} =
7.74^{+0.67}_{-0.57}~{\rm (stat.+syst.)~pb.} $$ The cross section
measurement is in good agreement with the SM expectation, and the $R$
measurement is within approximately 2.5 standard deviations of the SM
expectation, being the most precise measurement to date.

\section{Summary}
D0 is pursuing a wealth of top quark physics analyses that probe the
validity of the SM, a few of which are presented in this report. So
far, no evidence for new physics phenomena has been found in top quark
decays. The search continues now both at the Tevatron and the LHC, and
will include updates to the analyses presented here.


\section*{Acknowledgements}
The author would like to thank the organisers for the broad bandwidth
of topics addressed and the fruitful collaborative atmosphere at PANIC
2011 and the D0 collaboration and the staffs at Fermilab and
collaborating institutions.

\bibliographystyle{utphys.bst}
\bibliography{Panic11_Pleier}

\end{document}